\documentclass[aps,preprint,showpacs,superscriptaddress,groupedaddress]{revtex4}  
\usepackage{graphicx}  
\usepackage{dcolumn}   
\usepackage{bm}        
\usepackage{amssymb}   

\usepackage[colorlinks=true
,urlcolor=blue
,anchorcolor=blue
,citecolor=blue
,filecolor=blue
,linkcolor=black
,menucolor=blue
,pagecolor=blue
,linktocpage=true
]{hyperref}

\newcommand{\SO}[1]{\ensuremath{\mathrm{SO}(#1)}}
\newcommand{\SU}[1]{\ensuremath{\mathrm{SU}(#1)}}

\begin{document}
\preprint{OHSTPY-TH-13-001}

\title{Yukawa Unification Predictions with effective ``Mirage" Mediation}

\author{Archana Anandakrishnan}
\author{Stuart Raby}
\affiliation{The Ohio State University, 191 W. Woodruff Ave., Columbus, OH
43210}

\date{\today}

\begin{abstract}
In this letter we analyze the consequences, for the LHC, of gauge and third
family Yukawa coupling
unification with a particular set of boundary conditions defined at the GUT
scale, which we characterize as effective
``mirage" mediation.  We perform a global $\chi^2$ analysis including the
observables $M_W, M_Z, G_F, \alpha_{em}^{-1},$ $\alpha_s(M_Z), M_t,
m_b(m_b), M_\tau, BR(B \rightarrow X_s \gamma), BR(B_s \rightarrow \mu^+ \mu^-)$
and $M_{h}$.  The fit is performed in
the MSSM in terms of 10 GUT scale parameters, while $\tan\beta$ and $\mu$ are
fixed at the weak scale.  We find good fits to
the low energy data and a SUSY spectrum which is dramatically different than
previously studied in the context of Yukawa unification.
\end{abstract}


\maketitle

Gauge coupling unification in supersymmetric grand unified theories (SUSY GUTs)
\cite{Dimopoulos:1981yj,
Dimopoulos:1981zb, Ibanez:1981yh, Sakai:1981gr, Einhorn:1981sx, Marciano:1981un}
 provides an experimental hint for
low energy SUSY.   However, it does not significantly constrain the spectrum of
supersymmetric particles.  On the
other hand, it has been observed that Yukawa coupling unification for the third
generation of quarks and leptons
in models, such as $\SO{10}$ or $\SU{4}_c \times \SU{2}_L \times \SU{2}_R$, can
place significant constraints on the SUSY spectrum in order
to fit the top, bottom and tau masses
\cite{Blazek:2001sb,Baer:2001yy,Blazek:2002ta,Tobe:2003bc,Auto:2003ys}.   These
constraints depend on the particular boundary conditions for sparticle masses
chosen at the GUT scale (see for example,
\cite{Blazek:2002ta,Baer:2009ie,Badziak:2011wm,Anandakrishnan:2012tj},  which
consider different GUT scale boundary conditions).
In this letter we consider {\em effective ``mirage"} mediation boundary
conditions and show that they are consistent with gauge and Yukawa coupling
unification with a dramatically different low energy SUSY spectrum.  The GUT
scale boundary conditions are given by an  effective ``mirage" pattern with
gaugino masses defined in terms of two parameters, $M_{1/2}$ an overall mass
scale and $\alpha$ the ratio of the anomaly mediation to gravity mediation
contribution \cite{Choi:2005ge,Choi:2005hd,Kitano:2005wc,Lowen:2008fm}. Scalar masses are
given in terms of $m_{16}$ (for squarks and sleptons) and $m_{10}$ (for Higgs
doublets).  In addition, the $H_u$ and $H_d$ masses are split, either with
``Just-So" splitting or with a U(1) D-term which affects all scalar masses.
Note, as in Ref. \cite{Lowen:2008fm}, we allow for several origins of SUSY
breaking.  For example, the dilaton and conformal compensator fields break SUSY
at a scale of order $M_{1/2}$, while the dominant contribution to SUSY breaking
is at a scale of order $m_{3/2} \geq m_{16} \approx m_{10}$.  We fit the low
energy observables,  $M_W, M_Z, G_F, \alpha_{em}^{-1},$ $\alpha_s(M_Z), M_t,
m_b(m_b), M_\tau, BR(B \rightarrow X_s \gamma),  BR(B_s \rightarrow \mu^+
\mu^-)$ and $M_{h}$ in terms of 12 arbitrary parameters.   The low energy
sparticle spectrum is imminently amenable to testing at the LHC.  Two benchmark
points are contained in Table \ref{spectrum}.

Fermion masses and quark mixing angles are manifestly hierarchical.   The
simplest way to describe this hierarchy is with Yukawa matrices which
are also hierarchical.   Moreover the most natural way to obtain the hierarchy
is in terms of effective higher dimension operators of the form
\begin{equation}  W \supset \lambda \ 16_3 \ 10 \ 16_3 + 16_3 \ 10 \
\frac{45}{M} \ 16_2 + \cdots .
\end{equation}
This version of \SO{10} models has the nice features that it only requires small
representations of \SO{10},  has many predictions
and can, in principle, find an UV completion in string theory. The only
renormalizable term in $W$ is $\lambda \ 16_3 \ 10 \ 16_3$ which gives Yukawa
coupling unification
\begin{equation}  \lambda = \lambda_t = \lambda_b = \lambda_\tau =
\lambda_{\nu_\tau}  \end{equation} at $M_{GUT}$.
Note,  one {\it cannot} predict the top mass due to large SUSY threshold
corrections to the bottom and tau masses, as shown in
\cite{Hall:1993gn,Carena:1994bv,Blazek:1995nv}.  These corrections are of the
form
\begin{equation}  \delta m_b/m_b  \propto \frac{\alpha_3 \ \mu \ M_{\tilde g} \
\tan\beta}{m_{\tilde b}^2} +
\frac{\lambda_t^2 \ \mu \ A_t \ \tan\beta}{m_{\tilde t}^2} + {\rm log \
corrections} .
\end{equation} So instead  we use  Yukawa unification to predict the soft SUSY
breaking masses. In order to fit the data,
we need \begin{equation} \delta m_b/m_b \sim - 2\% . \end{equation}  We take
$\mu < 0$, $M_{\tilde g} > 0$.  For a short list of references on this subject,
see
\cite{Blazek:2001sb,Blazek:2002ta,Baer:2001yy,Auto:2003ys,Tobe:2003bc,
Dermisek:2003vn,Dermisek:2005sw,Baer:2008jn,Baer:2008xc,Gogoladze:2011aa,Gogoladze:2011ce}.

We assume the following GUT scale boundary conditions, namely a universal squark
and slepton mass parameter,  $m_{16}$,  universal cubic scalar parameter, $A_0$,
``mirage" mediation gaugino masses, \begin{equation} M_i = \left(1 + \frac{g_G^2
b_i \alpha}{16 \pi^2} \log \left(\frac{M_{Pl}}{m_{16}} \right) \right) M_{1/2}
\end{equation} (where $M_{1/2}$ and $\alpha$ are free parameters and $b_i =
(33/5, 1, -3) \; {\rm for} \; i = 1, 2, 3)$.  Note, this expression is equivalent to
the gaugino masses defined in \cite{Choi:2007ka}. $\alpha$ in the above expression is related to the
$\rho$ in Ref.\cite{Lowen:2008fm} as: $\frac{1}{\rho} = \frac{\alpha}{16 \pi ^2} {\rm ln} \frac{M_{PL}}{m_{16}}$.
We consider two different cases
for non-universal Higgs masses [NUHM] with ``just so'' Higgs splitting
\begin{equation} m_{H_{u (d)}}^2 = m_{10}^2  - (+) 2 D \end{equation} or, D-term
Higgs splitting, where, in addition, squark and slepton masses are given by
\begin{equation} m_a^2 =  m_{16}^2 + Q_a D, \;\; \{ Q_a = +1,  \{ Q, \bar u,
\bar e \}; -3, \{ L, \bar d \} \} \end{equation} with the U(1) D-term, $D$, and
SU(5) invariant charges, $Q_a$.   Note, we take $\mu, \, M_{1/2} < 0$. Thus for
$\alpha \ge 4 $ we have $M_3 > 0, M_1, M_2 < 0$.  (Note, the case of D-term
splitting is similar to the analysis of Ref.~\cite{Badziak:2011wm}. However our
low energy SUSY spectrum is much different.)  In the set of boundary
conditions above, the scalar masses and tri-linear couplings are large (of order $m_{3/2}$), while the magnitude of the gaugino masses
is given by $M_{1/2} \ll m_{3/2}$.  Note, this does not agree with the examples of mirage mediation in the literature.
For example, in the context of Type IIB strings, Ref. \cite{Choi:2005ge,Choi:2005hd,Kitano:2005wc}, the scalar, gaugino and tri-linear couplings
are all of order $m_{3/2}$, while in the heterotic version of mirage mediation, Ref. \cite{Lowen:2008fm}, the soft terms
for scalar masses are of order $m_{3/2}$, while the gaugino masses and tri-linear couplings are given by $M_{1/2} \ll m_{3/2}$.
Finding a SUSY breaking mechanism with the set of boundary conditions presented here is still an open challenge.  Nevertheless, we are using the SO(10) symmetry to justify Yukawa unification for the third family and then finding the minimal set of SUSY breaking parameters at the GUT scale
consistent with the low energy data.  This forces $A_0$ to be large.

\begin{figure}
\includegraphics[width=9cm]{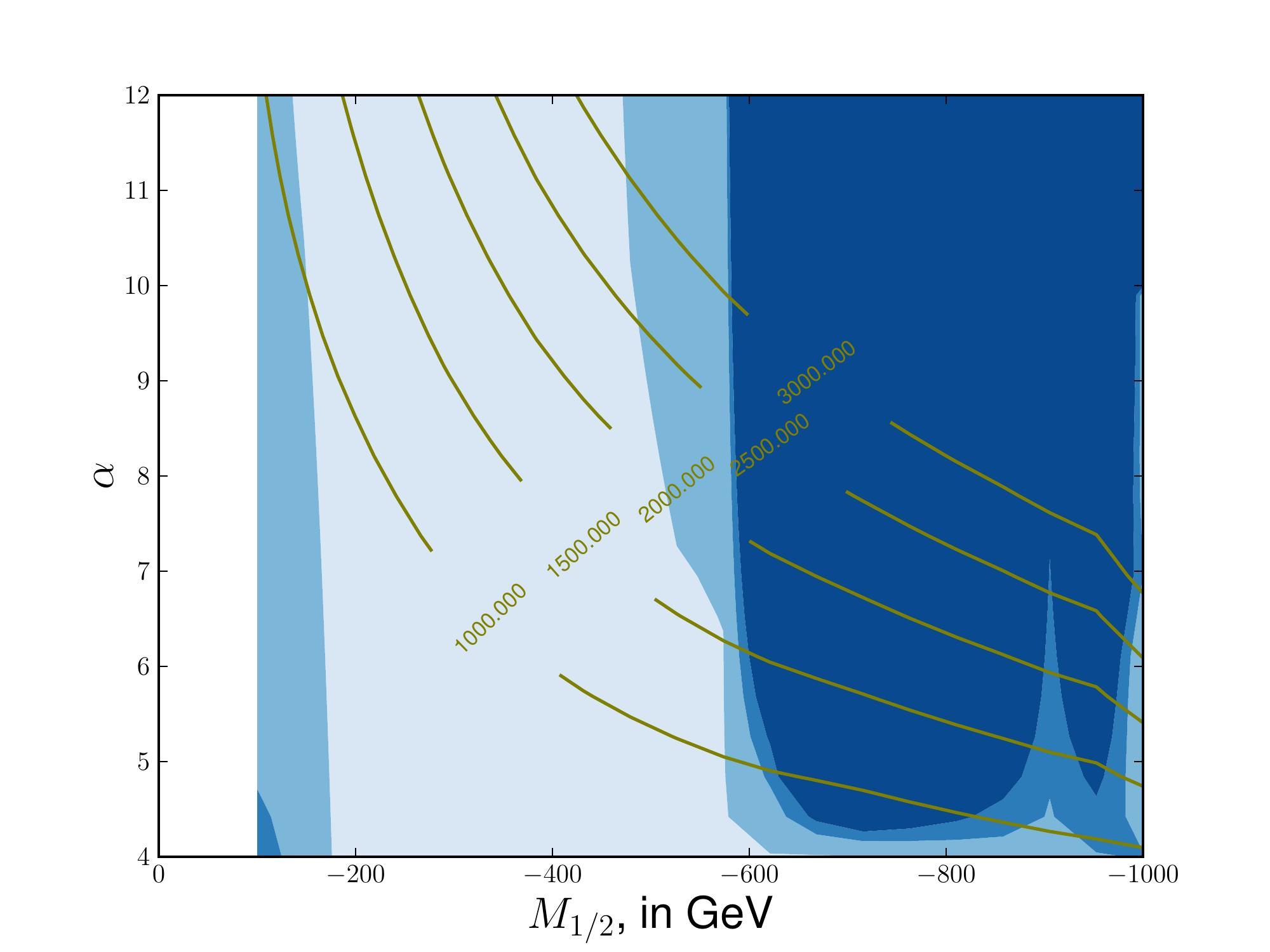}
\caption{\footnotesize The figure shows total $\chi^2$ in the $\alpha - M_{1/2}$ plane. The different shades
of blue regions have $\chi^2$/d.o.f = 1, 2.3, 3 and greater (from light to dark), and the olive curves show contours of constant gluino mass. }
\label{gluinochi}
\end{figure}

We perform a global $\chi^2$ analysis varying the parameters in  Table
\ref{tab:parameters} used to calculate the total $\chi^2$ function in terms
of all the observables given in  Table \ref{tab:allobservables} defined at the
electroweak scale as discussed in Ref. \cite{Anandakrishnan:2012tj}.
We minimize the $\chi^2$ function using the Minuit package maintained by CERN
\cite{James:1975dr}. Note that Minuit is not guaranteed to find the
\textit{global} minimum, but will in most cases converge on a local one. For
that reason, we iterate $\mathcal{O}$(100) times the minimization procedure for
each set of input parameters, and in each step we take a different initial guess
for the minimum (required by Minuit) so that we have a fair chance of finding
the true minimum.  We realize that the system is under-constrained and thus
we obtain values of $\chi^2 \ll 1$.  For this reason, it is not possible to define
a \textit{goodness of fit} or $\chi^2$/d.o.f.  However, in Fig. \ref{gluinochi},
we fix certain parameters such that we have 2 degrees of freedom, and plot contours of $\chi^2/dof = 1, 2.3, 3$
corresponding to 95\%, 90\%, and 68\% CLs, respectively. One could also add
more observables to the fit and this is possible when one considers
a three family model, which is the subject of an ongoing study.  The additional parameters determining
fermion masses, mixing angles and flavor observables for the first two families introduce more degrees of freedom (as discussed
previously in Ref. \cite{Anandakrishnan:2012tj} with different GUT scale boundary conditions), but they
do not significantly affect the SUSY spectrum.

\begin{table}
\begin{center}
\renewcommand{\arraystretch}{1.2}
\scalebox{0.83}{
\begin{tabular}{|l||c|}
\hline
Sector &  Third Family Analysis  \\
\hline
gauge             & $\alpha_G$, $M_G$, $\epsilon_3$          \\
SUSY (GUT scale)  & $m_{16}$, $M_{1/2}$, $\alpha$, $A_0$, $m_{10}$, $D$
\\
textures          & $\lambda$                                         \\
SUSY (EW scale)   & $\tan \beta$, $\mu$                               \\
\hline
Total \#          &                                              12 \\
\hline
\end{tabular}
}
\caption{\footnotesize The model is defined by three gauge parameters,
$\alpha_{G}, M_{G}$ (where $\alpha_1(M_G) = \alpha_2(M_G) \equiv \alpha_G)$, and $\epsilon_3 = \frac{\alpha_3 - \alpha_G}{\alpha_G}$; one large Yukawa coupling, $\lambda$; 6 SUSY
parameters defined at the GUT scale, $m_{16}$ (universal scalar mass for squarks
and sleptons), $M_{1/2}$ (universal gaugino mass), $\alpha$ (the ratio of
anomaly mediation to gravity mediation contribution to gaugino masses),
$m_{10}$, (universal Higgs mass), $A_0$ (universal trilinear scalar coupling)
and $D$ which fixes the magnitude of Higgs splitting in the case of ``Just-so"
Higgs splitting or the magnitude of all scalar splitting in the case of D-term
splitting.  The parameters $\mu, \ \tan\beta$ are obtained at the weak scale by
consistent electroweak symmetry breaking.}
\label{tab:parameters}
\end{center}
\end{table}

\begin{table}
\begin{center}
\begin{tabular}{|l|l|l|}
\hline
\textbf{Observable} &  \textbf{Exp.~Value}   & \textbf{Ref.}   \\
\hline
\hline
$\alpha_3(M_Z)$          &  $0.1184\pm0.0007$               &
\cite{Beringer:1900zz}      \\
$\alpha_{\text{em}}$  &  $1/137.035999074(44)$            &
\cite{Beringer:1900zz}      \\
$G_\mu$             &  $1.16637876(7)\times10^{-5}\text{ GeV}^{-2}$ &
\cite{Beringer:1900zz}      \\
$M_W$               &  $80.385\pm0.015\text{ GeV}$                     &
\cite{Beringer:1900zz}    \\
$M_Z$              &   $91.1876 \pm 0.0021$                             &
\cite{Beringer:1900zz}           \\
\hline
$M_t$               &  $173.5\pm1.0\text{ GeV}$                &
\cite{Beringer:1900zz}      \\
$m_b(m_b)$               &  $4.18\pm0.03\text{ GeV}$ & \cite{Beringer:1900zz}
  \\
$M_\tau$            &  $1776.82\pm0.16\text{ MeV}$                &
\cite{Beringer:1900zz}      \\
\hline
$M_h$               &  $125.3\pm0.4\pm0.5\text{ GeV}$                &
\cite{:2012gu}      \\
\hline
$\text{BR}(b \rightarrow s \gamma)$                &  $(343\pm21\pm7) \times
10^{-6}$      & \cite{hfag:2012-10-24}     \\
$\text{BR}(B_s  \rightarrow \mu^+ \mu^-) $          &  $(3.2 \pm 1.5)\times 10^{-9}$
& \cite{:2012ct}      \\
\hline
\end{tabular}
\end{center}
\caption{\footnotesize The 11 observables that we fit and their experimental
values.  Capital letters denote pole masses. We take LHCb results into account,
but use the average by Ref.~\cite{hfag:2012-10-24}. All experimental errors are
$1\sigma$ unless otherwise indicated.  Finally, the $Z$ mass is fit precisely
via a separate $\chi^2$ function solely imposing electroweak symmetry breaking.
}
\label{tab:allobservables}
\end{table}

Consider first the SUSY spectrum in our analysis.  Two benchmark points are
given in Table \ref{spectrum} with fixed $m_{16} = 5$ TeV.  The first and second
family squarks and sleptons have mass of order $m_{16}$, while
stops, sbottoms and staus are all a factor of about 2 lighter.  In addition,
gluinos are always lighter than the third family squarks
and sleptons, and the lightest charginos and neutralinos are even lighter.
Fig.\ref{gluinochi} shows that the gluino mass increases as
$\alpha$ increases and we are able to find good fits for gluino masses up to at least 3 TeV.
In models with universal gaugino masses, however, it was found that for fixed values of $m_{16}$, there is
an upper bound on the gluino mass \cite{Anandakrishnan:2012tj}, which is not the case here.
Note, CMS and ATLAS have used simplified models to place lower bounds on the
gluino mass.  However the allowed decay modes for our model, as presented below,
do not in any way resemble any simplified model.  Preliminary analysis,
Ref. \cite{anandakrishnan:2013ar},  shows that with such decay branching fractions
the bounds coming from published LHC data are at least 20\% lower than obtained
using any simplified model.  The states $\tilde{ \chi}^{\pm}_1$ and $\tilde{
\chi}^0_1$ are approximately degenerate.  In Table \ref{spectrum} we include the running masses
for the chargino and neutralino and the dominant one-loop contribution to the
mass splitting, $\Delta M$ \cite{Cheng:1998hc}.  Thus the chargino signature at
the LHC is dominated by the decay  $\tilde \chi^+ \rightarrow \tilde \chi^0 +
\pi^+$ \cite{Chen:1999yf}. This typically results in a disappearing charged
track since the pion would carry too little energy.  The present limits from
ATLAS are {\em not} very constraining \cite{ATLAS:2012jp}.  Our LSP is a wino-like
neutralino.   As a result, the thermal abundance of the LSP (obtained using
micrOMEGA 2.4 \cite{Belanger:2010pz}) is of order $10^{-5}$ due to the large
annihilation cross-section to $W^+ \, W^-$, i.e. too small for dark matter.
However, non-thermal production of wino dark matter can give the correct
abundance \cite{Moroi:1999zb,Gelmini:2006pw,Acharya:2008bk,Baer:2010kd}.
Finally, for the two benchmark points, the dominant decay modes for the gluino
are (calculated using Sdecay \cite{Muhlleitner:2003vg})
\begin{itemize}
\item for ``Just-so" Higgs splitting - (63\% $\rightarrow \tilde{\chi}^0 g$ ;
28\% $\rightarrow \tilde{\chi}^+  b \bar t + \tilde{\chi}^-  t
\bar b$
and 8\% $\rightarrow \tilde{\chi}^0  t  \bar t$)
\item and for D-term splitting -
(76\% $\rightarrow \tilde{\chi}^+  b  \bar t + \tilde{\chi}^-  t
\bar b$;
14\% $\rightarrow \tilde{\chi}^0  t  \bar t$;  3.5\% $\rightarrow \tilde{\chi}^0
 b  \bar b$,
and the rest to light quarks or gluons).
\end{itemize}
Note, in the case of  ``Just-so" Higgs splitting, stops are the lightest
sfermion, while in the case of D-term splitting, sbottoms are lighter.  In
addition, at low energies,  $A_t, A_b$ are small and thus we have small
left-right mixing.  These affect the gluino decay branching ratios.
Finally, since both $\mu, M_2$ and $M_1$ are negative we obtain the
correct sign for the SUSY correction to $(g - 2)_\mu$, however, in practice, our sleptons are too heavy to
give a good fit, and therefore $(g - 2)_\mu$ is not included in the $\chi^2$ function.  This does not agree with the
results of Badziak et al., Ref. \cite{Badziak:2011wm} who are able to fit $(g - 2)_\mu$, with non-universal gaugino masses
and Yukawa unification.
Unfortunately the sparticle spectrum obtained in their paper is now ruled out by LHC Higgs data \cite{Badziak:2012mm}.  In Tables \ref{justso} and
\ref{dterm} we give different benchmark points, all with $\chi^2 \ll 1$, in
order to present the variation of sparticle masses with different values of
$m_{16}$ and $M_{1/2}$.

With regards to GUT scale parameters, we find $\alpha \approx 12$  which
corresponds to approximately equal dilaton and anomaly mediated contributions to
gaugino masses.   We also find $|\epsilon_3| \leq 1\%$ in the case of D-term
splitting or precise gauge coupling unification \cite{Raby:2009sf}.

\begin{table}
\centering
\begin{tabular}{|l|c|c|}
\hline
{\rm NUHM}  & {\rm ``Just-so"} &  {\rm D-term} \\
\hline
\hline
$m_{16}$  &  5000  &  5000  \\
$\sqrt{D}$  & 1877  & 1242  \\
$m_{10}$  & 6097  &  5261 \\
$A_0$  & 8074  & 5593   \\
$\mu  $  & -615  &  -1294   \\
$M_{1/2}$ & -105   & -100   \\
$\alpha$  &  11.59   &  12.00   \\
$M_{GUT} \times 10^{-16}$   & 4.50     &  2.38   \\
$1/\alpha_{GUT}$ &  25.11   &  25.64  \\
$\epsilon_3$  &  -0.039  & -0.007  \\
$\lambda$ & 0.59  & 0.56  \\
$\tan\beta$ &  49.43 &  48.73 \\
\hline
$M_A$ & 1558 & 1237   \\
$m_{\tilde t_1}$ & 1975 & 2921    \\
$m_{\tilde b_1}$  & 2049 & 2159   \\
$m_{\tilde \tau_1}$  & 2473 & 3601    \\
$m_{\tilde u}$ & 4905 & 5081    \\
$m_{\tilde d}$  & 4944 & 4467   \\
$m_{\tilde e}$  & 4947 & 4477    \\
$m_{\tilde\chi^0_1}$   & 231.98 & 219.11      \\
$m_{\tilde\chi^+_1}$  & 232.05 & 219.11     \\
$\Delta M \equiv M_{\tilde \chi^+}- M_{\tilde \chi^0}$ & 0.519  &  0.438  \\
$M_{\tilde g}$   & 882 & 874     \\
\hline
\end{tabular}
\caption{\footnotesize Benchmark points and SUSY Spectrum. For each case we have
$\chi^2 \ll 1$.  The chargino and neutralino masses
are tree level and the one loop correction to the mass difference is given by
$\Delta M$.  All masses are in GeV.} \label{spectrum}
\end{table}

\begin{table}
\begin{tabular}{|l|c|c|c|c|}
\hline
$m_{16}$  &  4000  &  4000 & 10000 & 8000\\
$\sqrt{D}$  &1725  &1511 & 5516 & 3207 \\
$m_{10}$  & 5144  & 5079  &13036 & 10168 \\
$A_0$  & 7050  & 7542 & 15789 &  14687\\
$\mu  $  & -259 & -391  & -1364 & -612\\
$M_{1/2}$ & -100 & -240 & -120 & -260\\
$\alpha$  &   12.00  &  11.99  & 10.88 & 11.58 \\
$M_{GUT} \times 10^{-16}$ & 2.69 & 2.27 & 2.52 & 2.55  \\
$1/\alpha_{GUT}$ & 25.29 & 25.53 & 25.88 & 25.76 \\
$\epsilon_3$  & -0.019 & -0.017 & -0.005 & -0.018  \\
$\lambda$ & 0.616 & 0.616 & 0.560 & 0.606  \\
$\tan\beta$ &  50.25  &  49.96 & 48.68   & 49.93    \\
\hline
$M_A$ & 1658 & 1041 &6975 & 2825 \\
$m_{\tilde t_1}$ & 1308 & 1679 & 4028 & 2751  \\
$m_{\tilde b_1}$  & 1279 & 1760  & 3068& 2861 \\
$m_{\tilde \tau_1}$  & 1613 & 1580  & 5021 & 3282 \\
$m_{\tilde u}$ & 3929 & 4144  & 9659 & 7910 \\
$m_{\tilde d}$  & 3974 & 4155  & 9876 & 7978 \\
$m_{\tilde e}$  & 3952 & 3995   & 9808 & 7924 \\
$m_{\tilde\chi^0_1}$   & 187 & 367 & 278 &  525   \\
$m_{\tilde\chi^+_1}$  & 190 & 371   & 278&  526 \\
$\Delta M $ & 3.61 &  4.54 & 0.452 & 1.67 \\
$M_{\tilde g}$   & 858 & 1834  & 853 &  1902 \\
\hline
\end{tabular}
\caption{Generic Features of the ``Just-so" Higgs splitting with the mirage
pattern for gaugino masses and with different values of $m_{16}$ and $M_{1/2}$.
For each case we have $\chi^2 \ll 1$.  The chargino and neutralino masses
are tree level and the one loop correction to the mass difference is given by
$\Delta M$. All masses are in GeV.} \label{justso}
\end{table}

\begin{table}
\begin{tabular}{|l|c|c|c|c|}
\hline
$m_{16}$  &  4000 & 4000  & 8000 &8000\\
$\sqrt{D}$  & 1037  & 1018 & 2531 & 1641 \\
$m_{10}$  & 4598  & 4594 & 8094 & 7351 \\
$A_0$  & 5588  & 5654 & 8325 &  4810 \\
$\mu  $  & -541 & -591  & -2945 & -2636 \\
$M_{1/2}$ & -100  & -280 & -100 & -280 \\
$\alpha$  &  12.00   & 11.91  & 12.00 & 10.39 \\
$M_{GUT} \times 10^{-16}$ & 2.41 & 1.87 & 1.93 & 2.35 \\
$1/\alpha_{GUT}$ & 25.46 & 25.73 & 26.05 & 26.00 \\
$\epsilon_3$  & -0.011 & -0.009 & 0.007 & -0.009 \\
$\lambda$ & 0.582 & 0.599 & 0.540 & 0.569  \\
$\tan\beta$ &  49.20    &  49.31    & 48.13  &  48.70 \\
\hline
$M_A$ & 969 & 728 & 3719 & 726 \\
$m_{\tilde t_1}$ & 2026 & 2421 & 5178 & 5344  \\
$m_{\tilde b_1}$  & 1255 & 1825  & 2634 & 4702 \\
$m_{\tilde \tau_1}$  & 2622 & 2644  & 5266 & 6218 \\
$m_{\tilde u}$ & 4091 & 4324  & 8233 & 8105 \\
$m_{\tilde d}$  & 3553 & 3825  & 6594 & 7445 \\
$m_{\tilde e}$  & 3546 & 3615   & 6607 & 7445 \\
$m_{\tilde\chi^0_1}$   & 215 & 529 & 226 &    529 \\
$m_{\tilde\chi^+_1}$  & 216 & 531   & 226 &  529 \\
$\Delta M $ & 0.554 &  2.25 & 0.436 & 0.475 \\
$M_{\tilde g}$   & 867 & 2085  & 855 &  1842 \\
\hline
\end{tabular}
\caption{Generic Features of D-term Higgs splitting with the mirage pattern for
gaugino masses and with different values of $m_{16}$ and $M_{1/2}$.  For each
case we have $\chi^2 \ll 1$.  The chargino and neutralino masses
are tree level and the one loop correction to the mass difference is given by
$\Delta M$. All masses are in GeV.} \label{dterm}
\end{table}

In conclusion, we have performed a global $\chi^2$ analysis of an \SO{10} SUSY
GUT with gauge coupling
unification and top, bottom, $\tau$, $\nu_\tau$ Yukawa unification at $M_{GUT}$.
We have analyzed the model for the third family alone.  We have shown that the
SUSY spectrum is predominantly determined by fitting the third family and light
Higgs masses and the branching ratio $BR(B_s \rightarrow \mu^+ \ \mu^-)$.

A generic prediction of third family Yukawa unification is that we have
$\tan\beta \approx 50$.   In addition, in order to fit the branching ratio $BR(B_s
\rightarrow \mu^+ \ \mu^-)$ we find the CP odd Higgs mass, $m_A \gg M_Z$.
Hence we are in the decoupling limit and the light Higgs is predicted to be
Standard Model-like. Our model, makes several additional predictions which are unique to the effective ``mirage"
mediation boundary conditions.

\begin{itemize}
\item The first and second family of squarks and sleptons obtain mass of order
$m_{16}$, while the third family scalars are naturally about a factor of 2
lighter.  Gluinos and the lightest chargino and neutralino are always lighter
than the third family squarks and sleptons. We also find that there is no upper bound on the
gluino mass
\item Our LSP is predominantly wino and thus assuming a thermal calculation of
the relic abundance, we find $\Omega_{\tilde \chi^0_1} \sim 10^{-5}$.
\item $\tilde{ \chi}^{\pm}_1$ and $\tilde{ \chi}^0_1$ are approximately
degenerate.  Thus the chargino signature at the LHC is predominantly due to the
decay  $\tilde \chi^+ \rightarrow \tilde \chi^0 + \pi^+$. This typically results
in a disappearing charged track since the pion would carry too little energy.
\item  For the two benchmark points, Table \ref{spectrum}, the dominant decay
modes for the gluino are for ``just so" Higgs splitting - (63\% $\rightarrow
\tilde{ \chi}^0 \, g$;
28\% $\rightarrow \tilde{ \chi}^+  b  \bar t$, $\rightarrow \tilde{ \chi}^-  t
\bar b$
and 8\% $\rightarrow \tilde{ \chi}^0  t  \bar t$) and D-term splitting -
(76\% $\rightarrow \tilde{ \chi}^+  b  \bar t$, $\rightarrow \tilde{ \chi}^-  t
\bar b$;
14\% $\rightarrow \tilde{ \chi}^0  t  \bar t$;  3.5\% $\rightarrow \tilde{
\chi}^0  b  \bar b$,
and the rest to light quarks or gluons).
\end{itemize}

{\bf Acknowledgements}

{\small A.A. and S.R. are partially supported by DOE grant DOE/ER/01545-897.
A.A. thanks the Laboratoire de Physique Subatomique et de Cosmologie, Grenoble,
France for their kind hospitality while working on this
project. Finally, A.A. and S.R. thank Radovan Dermisek for the code Maton which
has been adapted to do a major portion of the top-down analysis and the Ohio
Supercomputer Center for their resources.
We are also grateful to Marek Olechowski, Marcin Badziak, Ak\i{}n Wingerter, Charles Bryant,
Christopher Hill and Linda Carpenter for discussions.}

\bibliography{bibliography}

\bibliographystyle{utphys}

\end{document}